\documentclass[12pt]{article}

\usepackage{amsfonts}

\def\d#1#2{\frac{{\rm d}#1}{{\rm d}#2}}

\makeatletter

\advance \textheight by \topskip
\let\@xp=\expandafter
\let\@nx=\noexpand
\providecommand\numberwithin[2]{%
  \@ifundefined{c@#1}{\@nocounterr{#1}}{%
    \@ifundefined{c@#2}{\@nocounterr{#2}}{%
    \@addtoreset{#1}{#2}%
    \toks@\@xp\@xp\@xp{\csname the#1\endcsname}%
    \@xp\xdef\csname the#1\endcsname
      {\@xp\@nx\csname the#2\endcsname
       .\the\toks@}}}}
\def\eqnarray{%
   \stepcounter{equation}%
   \def\@currentlabel{\p@equation\theequation}%
   \global\@eqnswtrue
   \m@th
   \global\@eqcnt\z@
   \tabskip\@centering
   \let\\\@eqncr
   $$\everycr{}\halign to\displaywidth\bgroup
       \hskip\@centering$\displaystyle\tabskip\z@skip{##}$\@eqnsel
      &\global\@eqcnt\@ne\hfil$\displaystyle{\hbox{}##\hbox{}}$\hfil
      &\global\@eqcnt\tw@ $\displaystyle{##}$\hfil\tabskip\@centering
      &\global\@eqcnt\thr@@ \hb@xt@\z@\bgroup\hss##\egroup
         \tabskip\z@skip
      \cr
}
\def\lefteqn#1{\hbox to 2em{$\displaystyle #1$\hss}}
\makeatother

\numberwithin{equation}{section}

\usepackage{amsfonts}

\begin{document}

\begin{titlepage}

\hbox to \hsize{\hfil math-ph/9807017}
\hbox to \hsize{\hfil IFT-P.047/98}
\hbox to \hsize{\hfil July, 1998}

\vfill

\Large \bf

\begin{center}
Riccati-type equations, \\ generalised WZNW equations, \\
and multidimensional Toda systems
\end{center}

\normalsize \rm

\vskip 0.2in

\begin{center}
L. A. Ferreira$^\ast$, J. F. Gomes$^\ast$, A. V.
Razumov$^\dagger$\\[0.05in]
M. V. Saveliev$^\ast$\footnote{On leave of absence from the
Institute for Hight Energy Physics, 142284 Protvino, Moscow Region,
Russia, saveliev@mx.ihep.su}, A. H. Zimerman$^\ast$\\[0.2in]
{\footnotesize $^\ast$Instituto de F\'\i sica Te\'orica - IFT/UNESP}
\\
{\footnotesize Rua Pamplona 145, 01405-900, S\~ao Paulo - SP,
Brazil}\\
{\footnotesize laf@ift.unesp.br, jfg@ift.unesp.br,
saveliev@ift.unesp.br, ahz@ift.unesp.br} \\[0.2in]
{\footnotesize $^\dagger$Institute for High Energy Physics} \\
{\footnotesize 142284 Protvino, Moscow Region, Russia} \\
{\footnotesize razumov@mx.ihep.su} 
\end{center}

\vskip 0.2in

\begin{abstract}
We associate to an arbitrary $\mathbb Z$-gra\-da\-tion of the Lie
algebra of a Lie group a system of Riccati-type first order
differential equations. The particular cases under consideration are
the ordinary Riccati and the matrix Riccati equations. The
multidimensional extension of these equations is given. The
generalisation of the associated Redheffer--Reid differential systems
appears in a natural way. The connection between the Toda systems and
the Riccati-type equations in lower and higher dimensions is
established. Within this context the integrability problem for those
equations is studied. As an illustration, some examples of the
integrable multidimensional Riccati-type equations related to the
maximally nonabelian Toda systems are given.
\end{abstract}

\vfill

\end{titlepage}

\section{Introduction}

To the present time there is a great number of papers in mathematics 
and physics devoted to various aspects of the matrix differential 
Riccati equation proposed in twenties by Radon in the context of the 
Lagrange variational problem. In particular, this equation has been 
discussed in connection with the oscillation of the solutions to 
systems of linear differential equations, Lie group and differential 
geometry aspects of the theory of analytic functions of several
complex variables in classical domains, the probability theory,
computation schemes. For a systematic account of the development in
the theory of the matrix differential Riccati equation up to
seventies see, for example, the survey \cite{ZI73}. More recently
there appeared papers where this equation was considered as a
B\"acklund-type transformation for some integrable systems of
differential geometry, in particular, for the Lam\'e and the Bourlet
equations, and a relevant superposition principle for the equation
has been studied on the basis of the theory of Lie algebras, see, for
example, \cite{TT80} and references therein. The matrix Riccati
equation also arises as equation of motion on Grassmann manifolds and
on homogeneous spaces attached to the Hartree--Fock--Bogoliubov
problem, see, for example, \cite{DWO84BG92} and references  therein;
and in some other subjects of applied mathematics and physics such as
optimal control theory, plasma, etc., see, for example,
\cite{HM82Sch73Sh80DFM87}. Continued--fraction solutions to the
matrix differential Riccati equation were constructed
in \cite{CHR8690}, based on a sequence of substitutions with the
coefficients satisfying a matrix generalisation of 
the Volterra-type equations which in turn provide a B\"acklund 
transformation for the corresponding matrix version of the
Toda lattice. In papers \cite{B90} the matrix differential Riccati
equation occurs in the steepest descent solution to the total least
squares problem as a flow on Grassmannians via the Brockett double
bracket commutator equation; in the special case of projective space
this is the Toda lattice flow in Moser's variables. 

In the present paper we investigate the equations associated with an
arbitrary $\mathbb Z$-gradation of the Lie algebra $\mathfrak g$ of a
Lie group $G$. For the case $G = {\rm GL}(2, \mathbb C)$ and the
principal gradation of $\mathfrak{gl}(2, \mathbb C)$ this is the
ordinary Riccati equation, for the case $G = {\rm GL}(n, \mathbb C)$
and some special $\mathbb Z$-gradation of $\mathfrak{gl}(n, \mathbb
C)$ we get the matrix Riccati equation. The underlying
group-algebraic structure allows us to give a unifying approach to
the investigation of the integrability problem for the equations
under consideration which we call the Riccati-type equations.

We also give a multidimensional generalisation of the Riccati-type
equations and discuss their integrability.

It appeared very useful for the study of ordinary matrix Riccati
equations to associate with them the so-called Redheffer--Reid
differential system \cite{Red56Rei59}. In our approach the
corresponding generalisation of such systems appears in a natural
way. The associated Redheffer--Reid system can be considered as the
constraints providing some reduction of the
Wess--Zumino--Novikov--Witten (WZNW) equations. From the other hand,
it is well known that the Toda-type systems can be also obtained by
the appropriate reduction of the WZNW equations, see, for example,
\cite{FRRTW92}. This implies the deep connection of the Toda-type
systems and the Riccati-type equations. In particular, under the
relevant constraints the Riccati-type equations play the role of a
B\"acklund map for the Toda systems, and, in a sense,
are a generalisation of the Volterra equations.

Some years ago there appeared a remarkable generalisation \cite{GM93}
of the Wess--Zumino--Novikov--Witten (WZNW) equations.
The associated Redheffer--Reid system in the multidimensional case
can be considered again as the constraints imposed on the solutions
of those equations. We show that in the same way as in two
dimensional case, the appropriate reduction of the multidimensional
WZNW equations leads to the multidimensional Toda systems
\cite{RS97}, in particular to the equations \cite{CV91Dub93}
describing topological and antitopological fusion.\footnote{It is
rather clear that the the multidimensional systems suggested in
\cite{RS97} become two dimensional equations only under a relevant
reduction. Moreover, arbitrary mappings determining the general
solution to these equations are not necessarily factorised to the
products of mappings each depending on one coordinate only. One can
easily get convinced of it just by the examples considered there in
detail.} The multidimensional Toda systems are integrable for the
relevant integration data with the general solution being determined
by the corresponding arbitrary mappings in accordance with the 
integration scheme developed in \cite{RS97}. Therefore the
integrability problem for the multidimensional Riccati-type equations
can be studied, in particular, on the basis of that fact. As an
illustration of the general construction we discuss in detail some
examples related to the maximally nonabelian Toda systems
\cite{RS97a}.

Analogously to the Toda systems one can construct higher grading
generalisations in the sense of \cite{GS95,RS97b} for the
multidimensional Riccati-type equations.

\section{One dimensional Riccati--type equations}

Let $G$ be a connected Lie group and $\mathfrak g$ be its Lie
algebra. Without any loss of generality we assume that $G$ is a
matrix Lie group, otherwise we replace $G$ by its image under some
faithful representation of $G$. For any fixed mapping $\lambda:
\mathbb R \to \mathfrak g$ consider the equation 
\begin{equation}
\psi^{-1} \d \psi x = \lambda \label{2}
\end{equation}
for the mapping $\psi: \mathbb R \to G$. Certainly one can use the
complex plane $\mathbb C$ instead of the real line $\mathbb R$. 

Suppose that the Lie algebra $\mathfrak g$ is endowed with a $\mathbb
Z$-gradation,
\[
{\mathfrak g} = \bigoplus_{m \in {\mathbb Z}} \mathfrak g_m.
\]
Define the following nilpotent subalgebras of $\mathfrak g$:
\[
\mathfrak g_{<0} = \bigoplus_{m < 0} \mathfrak g_m, \qquad
\mathfrak g_{>0} = \bigoplus_{m > 0} \mathfrak g_m,
\]
and represent the mapping $\lambda$ in the form
\[
\lambda = \lambda_{<0} + \lambda_0 + \lambda_{>0},
\]
where the mappings $\lambda_{<0}$, $\lambda_0$ and $\lambda_{>0}$ 
take values in $\mathfrak g_{<0}$, $\mathfrak g_0$ and $\mathfrak
g_{>0}$ respectively. 

Denote by $G_{<0}$, $G_0$ and $G_{>0}$ the connected Lie subgroups of
$G$ corresponding to the subalgebras $\mathfrak
g_{<0}$, $\mathfrak g_0$ and $\mathfrak g_{>0}$ respectively. Under
the appropriate assumptions for an element $a \in G$ belonging to 
some dense subset of $G$ it is valid the generalised Gauss
decomposition
\begin{equation}
a = a_{<0} \, a_0 \, a_{>0}, \label{7}
\end{equation}
where $a_{<0} \in G_{<0}$, $a_0 \in G_0$ and $a_{>0} \in G_{>0}$. For
the mapping $\psi$ we can write
\begin{equation}
\psi = \psi_{<0} \, \psi_0 \, \psi_{>0}, \label{3}
\end{equation}
where the mapping $\psi_{<0}$ takes values in $G_{<0}$, the mapping
$\psi_0$ takes values in $G_0$ and the mapping $\psi_{>0}$ takes
values in $G_{>0}$. Using the Gauss decomposition (\ref{3}) of the
mapping $\psi$ rewrite equation (\ref{2}) as
\begin{equation}
\psi^{-1}_{>0} \, \left( \psi_{\le 0}^{-1} \, \d
{\psi_{\le 0}} x \right) \, \psi_{>0} + \psi^{-1}_{>0} \,
\d {\psi_{>0}} x = \lambda, \label{5} 
\end{equation}
where $\psi_{\le 0} = \psi_{<0} \psi_0$. From (\ref{5}) it follows
that
\[
\psi_{\le 0}^{-1} \, \d{\psi_{\le 0}} x +
\d{\psi_{>0}} x \, \psi^{-1}_{>0} = \psi_{>0} \, \lambda \,
\psi^{-1}_{>0},
\]
and hence
\begin{equation}
\psi_{\le 0}^{-1} \, \d{\psi_{\le 0}} x = (\psi_{>0}
\, \lambda \, \psi^{-1}_{>0})_{\le 0}, \label{6}
\end{equation}
where the subscript $\le 0$ denotes the corresponding component with
respect to the decomposition
\[
\mathfrak g = \mathfrak g_{\le 0} \oplus \mathfrak g_{>0} =
(\mathfrak g_{<0} \oplus \mathfrak g_0) \oplus \mathfrak g_{>0}.
\]
Substituting (\ref{6}) into (\ref{5}) one gets
\[
\psi^{-1}_{>0} \, \d{\psi_{>0}} x =
\lambda - \psi^{-1}_{>0} \, (\psi_{>0} \, \lambda \,
\psi^{-1}_{>0})_{\le 0} \, \psi_{>0}
\]
that can be rewritten as
\begin{equation}
\d{\psi_{>0}} x \, \psi^{-1}_{>0} = (\psi_{>0} \, \lambda \,
\psi^{-1}_{>0})_{>0}. \label{44}
\end{equation}
By the reasons which are clear from what follows we call this
equation for the mapping $\psi_{>0}$ a {\it Riccati-type
equation}.

The formal integration of equation (\ref{44}) can be performed in the
following way. Consider (\ref{2}) as a linear differential equation
for the mapping $\psi$:
\begin{equation}
\d \psi x = \psi \, \lambda. \label{4}
\end{equation}
Find the solution of this equation with the initial condition
$\psi(0) = a$, where $a$ is a constant element of the Lie group $G$.
Using now the Gauss decomposition (\ref{3}) of the mapping $\psi$
we find the solution of equation (\ref{44}) with the initial
condition
$\psi_{>0} = a_{>0}$, where $a_{>0}$ is the positive grade component
of $a$
arising from the Gauss decomposition (\ref{7}). It is clear that in
order to obtain the general solution of equation (\ref{44}) it
suffices to consider elements $a$ belonging to the Lie subgroup
$G_{>0}$. Then the solution of (\ref{44}) is expressed in terms of
solution of (\ref{4}). Note that the solution of equation (\ref{4})
with the initial condition $\psi(0) = a$ can be obtained from the
solution with the initial condition $\psi(0) = e$, where $e$ is
the unit element of $G$, by left multiplication by $a$.

Thus we have shown that one can associate a Ricatti-type equation
to any $\mathbb Z$-gradation of a Lie group. The integration of
this equations is reduced to integration of some matrix system of
first order linear differential equations.

Let now $\chi$ be some mapping from $\mathbb R$ to $G$. It is clear
that if the mapping $\psi$ satisfies equation (\ref{4}), then the
mapping $\psi' = \psi \chi^{-1}$ satisfies the equation
\[
\d {\psi'} x = \psi' \, \lambda',
\]
where
\begin{equation}
\lambda' = \chi \, \lambda \, \chi^{-1} - \d \chi x \, \chi^{-1}.
\label{9}
\end{equation}
If $\chi$ is a mapping from $\mathbb R$ to $G_0$, then 
the corresponding component
\[
\psi'_{>0} = \chi \, \psi_{>0} \, \chi^{-1}
\]
of the mapping $\psi'$ satisfies the Ricatti-type equation
(\ref{44}) with $\lambda$ replaced by $\lambda'$.
In this,
\[
\lambda'_0 = \chi \, \lambda_0 \, \chi^{-1} - \d \chi x \, \chi^{-1},
\]
and it is clear that we can choose the mapping $\chi$ so that
$\lambda'_0$ vanishes.

Another interesting possibility arises when $\chi$ is a mapping
from
$\mathbb R$ to $G_{>0}$. Let us choose a mapping $\chi$ such that
$\lambda'_{>0} = 0$. From (\ref{9}) it follows that this case is 
realised if and only if
\[
\d \chi x \, \chi^{-1} = (\chi \, \lambda \chi^{-1})_{>0},
\]
i.e., $\chi$ should satisfy the Riccati-type equation
(\ref{44}). Thus, having a particular solution of the Riccati-type
equation, its general solution can be constructed from the general
solution of the equation with $\lambda_{>0} = 0$. As will be shown
below, for this case the Riccati-type equation can be solved in a
quite simple way.

\section{Simplest example}
\label{Simplest}

Consider first the case of the Lie group GL$(n, \mathbb C)$, $n \geq
2$ and represent $n$ as the sum of two positive integers $n_1$ and
$n_2$. For the Lie algebra $\mathfrak{gl}(n, \mathbb C)$ there is a
$\mathbb Z$-gradation where arbitrary elements $x_{<0}$, $x_0$ and
$x_{>0}$ of the subalgebras $\mathfrak g_{<0}$, $\mathfrak g_{>0}$
and $\mathfrak g_0$ have the form
\[
x_{<0} = \left( \begin{array}{cc}
0 & 0 \\
(x_{<0})_{21} & 0
\end{array} \right), \qquad
x_0 = \left( \begin{array}{cc}
(x_0)_{11} & 0 \\
0 & (x_0)_{22}
\end{array} \right), \qquad
x_{>0} = \left( \begin{array}{cc}
0 & (x_{>0})_{12} \\
0 & 0
\end{array} \right).
\]
Here $(x_{<0})_{21}$ is an $n_2 \times n_1$ matrix, $(x_{>0})_{12}$
is an $n_1 \times n_2$ matrix, $(x_0)_{11}$ and $(x_0)_{22}$ are $n_1
\times n_1$ and $n_2 \times n_2$ matrices respectively. The
corresponding subgroups $G_{<0}$, $G_{>0}$ and $G_0$ are formed by
the matrices 
\[
a_{<0} = \left( \begin{array}{cc}
I_{n_1} & 0 \\
(a_{<0})_{21} & I_{n_2}
\end{array} \right), \qquad
a_0 = \left( \begin{array}{cc}
(a_0)_{11} & 0 \\
0 & (a_0)_{22}
\end{array} \right), \qquad
a_{>0} = \left( \begin{array}{cc}
I_{n_1} & (a_{>0})_{12} \\
0 & I_{n_2}
\end{array} \right).
\]
Here $(a_{<0})_{21}$ is an arbitrary $n_2 \times n_1$ matrix,
$(a_{>0})_{12}$ is an arbitrary $n_1 \times n_2$ matrix, $(a_0)_{11}$
and $(a_0)_{22}$ are arbitrary nondegenerate $n_1 \times n_1$ and
$n_2 \times n_2$ matrices respectively. The Gauss decomposition
(\ref{7}) of an element
\[
a = \left( \begin{array}{cc}
a_{11} & a_{12} \\
a_{21} & a_{22}
\end{array} \right)
\]
is given by the relations
\begin{eqnarray}
(a_0)_{11} = a_{11}, &\qquad& (a_{>0})_{12} = a_{11}^{-1} a_{12},
\label{10} \\
(a_{<0})_{21} = a_{21} a_{11}^{-1}, &\qquad& (a_0)_{22} = a_{22} -
a_{21} a_{11}^{-1} a_{12}. \label{11}
\end{eqnarray}

Parametrizing the mapping $\lambda$ as
\[
\lambda = \left( \begin{array}{cc}
A & B \\
C & D
\end{array} \right)
\]
and $\psi_{>0}$ as
\begin{equation}
\psi_{>0} = \left( \begin{array}{cc}
I_{n_1} & U \\
0 & I_{n_2} 
\end{array} \right), \label{41}
\end{equation}
one easily sees that equation (\ref{44}) takes in the case under
consideration the form
\begin{equation}
\d U x = B - AU + UD - UCU. \label{8}
\end{equation}
In the case $n=2$, $n_1 = n_2 = 1$, we have the usual Riccati
equation. For $n = 2m$, $n_1 = n_2 = m$, we come to the so-called
matrix Riccati equation. This justifies our choice for the name of
equation (\ref{44}) in general case.

\subsection{Case $B = 0$} 

If $C = 0$ then equation (\ref{8}) is linear. In the case
$B = 0$, under the conditions $n_1 = n_2$ and $\det U(x) \ne 0$
for any $x$, the substitution $V = U^{-1}$ leads to the linear
equation
\[
\d V x = V A - D V + C. 
\]
Nevertheless, it is instructive to consider the procedure of
obtaining  the general solution to equation (\ref{8}) for $B = 0$. 
Recall that having a particular solution to the Riccati-type
equation, we can reduce the consideration to the case where
$\lambda_{>0} = 0$. For the equation in question this is equivalent
to the requirement $B = 0$.

First, find the mapping $\chi: \mathbb R \to G_0$ such that
transformation (\ref{9}) would give $\lambda'_0 = 0$.
Parametrising $\chi$ as
\[
\chi = \left( \begin{array}{cc}
Q & 0 \\
0 & R 
\end{array} \right),
\]
one comes to the following equations for $R$ and $Q$:
\[
\d Q x = Q \, A, \qquad \d R x = R \, D.
\]
Therefore we can choose
\begin{equation}
Q(x) = P \exp \left( \int_0^x A(x') \, {\rm d}x' \right), \quad
R(x) = P \exp \left( \int_0^x D(x') \, {\rm d}x' \right),
\label{12}
\end{equation}
where the symbol $P \exp(\cdot)$ denotes the path
ordered exponential (multiplicative integral). Now solve the equation
\[
\d {\psi'} x = \psi' \lambda',
\]
where
\[
\lambda' = \left( \begin{array}{cc}
0 & 0 \\
C' & 0
\end{array} \right) =
\left( \begin{array}{cc}
0 & 0 \\
R C Q^{-1} & 0
\end{array} \right).
\]
The solution of this equation with the initial condition $\psi'(0)
= I_n$ is
\[
\psi(x) = 
\left( \begin{array}{cc}
I_{n_1} & 0 \\
S(x) & I_{n_2}
\end{array} \right)
\]
with
\begin{equation}
S(x) = \int_0^x R(x') \, C(x') \, Q^{-1}(x') \, {\rm d}x'. \label{13}
\end{equation}
Hence, the solution of equation (\ref{4}) with the initial
condition $\psi(0) = I_n$ is given by
\[
\psi = \left( \begin{array}{cc}
Q & 0 \\
S \, Q & R 
\end{array} \right).
\]
To obtain the general solution of the equation under consideration we
should have the solution of equation (\ref{4}) with the initial
condition
\begin{equation}
\psi(0) = \left( \begin{array}{cc}
I_{n_1} & m \\
0 & I_{n_2}
\end{array} \right), \label{16}
\end{equation}
where $m$ is an arbitrary $n_1 \times n_2$ matrix. Such a solution is
represented as
\[
\psi = \left( \begin{array}{cc}
(I_{n_1} + m S) Q & m R \\
S Q & R
\end{array} \right).
\]
Now, using (\ref{10}) we conclude that the general solution to
equation (\ref{8}) in the case $B = 0$ is
\[
U = Q^{-1} (I_{n_1} + m S)^{-1} m R,
\]
where $Q$, $R$ and $S$ are given by relations (\ref{12}) and
(\ref{13}).

Thus we see that in the case when $\lambda$ is a block upper or
lower triangular matrix the Riccati-type equation (\ref{8}) can be
explicitly integrated. Actually if $\lambda$ is a constant mapping we
can reduce it by a similarity transformation to the block upper or
lower triangular form and solve the corresponding Riccati-type
equation. The solution of the initial equation is obtained then by
some algebraic calculations.

\subsection{The case $A = 0$ and $D = 0$}

Representing the mapping $\psi$ in the form
\[
\psi = \left( \begin{array}{cc}
\psi_{11} & \psi_{12} \\
\psi_{21} & \psi_{22}
\end{array} \right)
\]
one easily sees that equation (\ref{4}) is equivalent to the system
\begin{eqnarray}
&\displaystyle \d {\psi_{11}} x = \psi_{12} C, \qquad \d
{\psi_{12}} x = \psi_{11} B, \label{14} \\
&\displaystyle \d {\psi_{21}} x = \psi_{22} C, \qquad \d
{\psi_{22}} x = \psi_{21} B. \label{15}
\end{eqnarray}

\subsubsection{The case $C = B$}

Consider the case $C = B$; that is certainly possible only if $n_1 =
n_2$. In this case we can rewrite equations (\ref{14}) and (\ref{15})
as
\begin{eqnarray*}
&\displaystyle \d {(\psi_{11} + \psi_{12})} x = (\psi_{11} +
\psi_{12} )B, \qquad \d {(\psi_{11} - \psi_{12})} x =
-(\psi_{11} - \psi_{12}) B, \\
&\displaystyle \d {(\psi_{22} + \psi_{21})} x = (\psi_{22} +
\psi_{21}) B, \qquad \d {(\psi_{22} - \psi_{21})} x = -
(\psi_{22} - \psi_{21}) B.
\end{eqnarray*}
Hence, the solution of equation (\ref{4}) with the initial condition
$\psi(0) = I_n$ is given by
\[
\psi = \frac{1}{2} \left( \begin{array}{cc}
F + H & F - H \\
F - H & F + H 
\end{array} \right),
\]
where
\[
F(x) = P \exp \left( \int_0^x B(x') \, {\rm d} x' \right), \quad H(x)
= P \exp \left(- \int_0^x B(x') \, {\rm d} x' \right).
\]
The solution of equation (\ref{4}) with the initial condition of form
(\ref{16}) is
\[
\psi = \frac{1}{2} \left( \begin{array}{cc}
F + H + m(F - H)& F - H + m(F + H)\\
F - H & F + H 
\end{array} \right);
\]
therefore, the general solution to the Riccati-type equation under
consideration can be written as
\[
U = (F + H + m(F - H))^{-1} (F - H + m(F + H)).
\]

\subsubsection{The case of constant $B$ and $C$}

As we noted above, the general solution to the Ricatti-type
equations (\ref{8}) for the case of constant mapping $\lambda$ can be
obtained by a reduction of $\lambda$ to the block upper or lower
triangular form. Nevertheless, it is interesting to consider the
particular case of constant $\lambda$ when the general solution has
the most simple form.

Suppose that $n_1 = n_2$ and that $B$ and $C$ are constant
nondegenerate matrices. In this case the solution of equation
(\ref{4}) with the initial condition $\psi(0) = I_n$ is
\[
\psi(x) = \left( \begin{array}{cc}
\cosh (\sqrt{BC} x) & \sinh(\sqrt{BC} x)\sqrt{BC} C^{-1} \\
\sinh (\sqrt{CB} x) \sqrt{CB} B^{-1} & \cosh (\sqrt{CB} x)
\end{array} \right),
\]
and for the general solution one has
\begin{eqnarray*}
U(x) = \Bigl( \cosh (\sqrt{BC} x) &+& m \sinh(\sqrt{CB} x)
\sqrt{CB} B^{-1} \Bigr)^{-1} \\
&\times& \Bigl( \sinh (\sqrt{BC} x) \sqrt{BC} C^{-1} + m
\cosh(\sqrt{CB} x) \Bigr).
\end{eqnarray*}
It should be noted here that the expression for $U(x)$
does not actually contain square roots of matrices that can be easily
seen from the corresponding expansions into the power series.

\section{A further example}

The next example is based on another $\mathbb Z$-gradation of the Lie
algebra $\mathfrak{gl}(n, \mathbb C)$. Here one represents $n$ as the
sum of three positive integers $n_1$, $n_2$ and $n_3$ and consider an
element $x$ of $\mathfrak{gl}(n, \mathbb C)$ as a $3 \times 3$ block
matrix $(x_{rs})$ with $x_{rs}$ being an $n_r \times n_s$ matrix. The
subspace $\mathfrak g_m$ is formed by the block matrices $x =
(x_{rs})$ where only the blocks $x_{rs}$ with $s - r = m$ are
different from zero. Arbitrary elements $x_{<0}$, $x_0$ and $x_{>0}$
of the subalgebras $\mathfrak g_{<0}$, $\mathfrak g_0$ and $\mathfrak
g_{>0}$ have the form
\begin{eqnarray*}
& x_{<0} = \left( \begin{array}{ccc}
0 & 0 & 0 \\
(x_{<0})_{21} & 0 & 0 \\
(x_{<0})_{31} & (x_{<0})_{32} & 0
\end{array} \right), \qquad
x_{>0} = \left( \begin{array}{ccc}
0 & (x_{>0})_{12} & (x_{>0})_{13} \\
0 & 0 & (x_{>0})_{23} \\
0 & 0 & 0
\end{array} \right), \\
& x_0 = \left( \begin{array}{ccc}
(x_0)_{11} & 0 & 0 \\
0 & (x_0)_{22} & 0 \\
0 & 0 & (x_0)_{33}
\end{array} \right).
\end{eqnarray*}
The subgroups $G_{<0}$, $G_0$ and $G_{>0}$ are formed by the
nondegenerate matrices
\begin{eqnarray*}
& a_{<0} = \left( \begin{array}{ccc}
I_{n_1} & 0 & 0 \\
(a_{<0})_{21} & I_{n_2} & 0 \\
(a_{<0})_{31} & (a_{<0})_{32} & I_{n_3}
\end{array} \right), \qquad
a_{>0} = \left( \begin{array}{ccc}
I_{n_1} & (a_{>0})_{12} & (a_{>0})_{13} \\
0 & I_{n_2} & (a_{>0})_{23} \\
0 & 0 & I_{n_3}
\end{array} \right), \\
& a_0 = \left( \begin{array}{ccc}
(a_0)_{11} & 0 & 0 \\
0 & (a_0)_{22} & 0 \\
0 & 0 & (a_0)_{33}
\end{array} \right).
\end{eqnarray*}
The Gauss decomposition of an element $a \in {\rm GL}(n, \mathbb C)$
is determined by the relations
\begin{eqnarray*}
&& (a_{<0})_{21} = a_{21} a_{11}^{-1}, \quad (a_{<0})_{31} = a_{31}
a_{11}^{-1}, \\
&& (a_{<0})_{32} = (a_{32} - a_{31} a_{11}^{-1} a_{12}) (a_{22} -
a_{21} a_{11}^{-1} a_{12})^{-1}, \\
&& (a_0)_{11} = a_{11}, \quad (a_0)_{22} = a_{22} - a_{21}
a_{11}^{-1} a_{12}, \\
&& (a_0)_{33} = a_{33} - a_{31} a_{11}^{-1} a_{13} \\
&& \hskip 4em {} - (a_{32} - a_{31}
a_{11}^{-1} a_{12})(a_{22} - a_{21} a_{11}^{-1} a_{12})^{-1} (a_{23}
- a_{21} a_{11}^{-1} a_{13}), \\
&& (a_{>0})_{12} = a_{11}^{-1} a_{12}, \quad (a_{>0})_{13} =
a_{11}^{-1} a_{13}, \\
&& (a_{>0})_{23} = (a_{22} - a_{21} a_{11}^{-1} a_{12})^{-1} (a_{23}
- a_{21} a_{11}^{-1} a_{13}).
\end{eqnarray*}

We parametrise the mapping $\lambda$ as
\[
\lambda = \left( \begin{array}{ccc}
A_{11} & B_{12} & B_{13} \\
C_{21} & A_{22} & B_{23} \\
C_{31} & C_{32} & A_{33}
\end{array} \right)
\]
and the mapping $\psi_{>0}$ as
\[
\psi_{>0} = \left( \begin{array}{ccc}
I_{n_1} & U_{12} & U_{13} \\
0 & I_{n_2} & U_{23} \\
0 & 0 & I_{n_3}
\end{array} \right).
\]
After some algebra one sees that the Riccati-type equations for
the case under consideration is
\begin{eqnarray*}
&& \d{U_{12}} x = B_{12} - A_{11}U_{12} + U_{12}A_{22} + U_{13}C_{32}
- U_{12}C_{21}U_{12} - U_{13}C_{31}U_{12}, \\
&& \d{U_{23}} x  = B_{23} - A_{22}U_{23} + U_{23}A_{33} -
C_{21}U_{13} \\
&& \hskip 6em {} + C_{21}U_{12}U_{23} - U_{23}C_{31}U_{13} -
U_{23}C_{32}U_{23} + U_{23}C_{31}U_{12}U_{23},\\
&& \d{U_{13}} x = B_{13} - A_{11}U_{13} + U_{13}A_{33}
+ U_{12}B_{23} - U_{12}C_{21}U_{13} - U_{13}C_{31}U_{13}. 
\end{eqnarray*}

Consider the case where $B_{rs} = 0$. Here by transformation
(\ref{9}) we can reduce our equations to the case
where additionally $A_{rs} = 0$. In the latter case the solution
of equation (\ref{4}) with the initial condition $\psi(0) = I_n$ has
the form
\[
\psi = \left( \begin{array}{ccc}
I_{n_1} & 0 & 0 \\
S_{21} & I_{n_2} & 0 \\
S_{31} & S_{32} & I_{n_3}
\end{array} \right),
\]
where
\begin{eqnarray*}
&& S_{21}(x) = \int_0^x C_{21}(x') \, {\rm d} x', \\
&& S_{31}(x) = \int_0^x \left( C_{31}(x') + \left( \int_0^{x'}
C_{32}(x'') {\rm d} x'' \right) C_{21}(x') \right) {\rm d} x', \\
&& S_{32}(x) = \int_0^x C_{32}(x') \, {\rm d} x'.
\end{eqnarray*}
Using the explicit expressions for the Gauss decomposition given in
this section, we find that the solution to the Riccati-type equation
under consideration with the initial condition
\[
\psi_{>0}(0) = \left( \begin{array}{ccc}
I_{n_1} & m_{12} & m_{13} \\
0 & I_{n_2} & m_{23} \\
0 & 0 & I_{n_3}
\end{array} \right)
\]
is determined by the relations
\begin{eqnarray*}
&& U_{12} = (I_{n_1} + m_{12} S_{21} + m_{13} S_{31})^{-1} (m_{12} +
m_{13} S_{32}), \\
&& U_{13} = (I_{n_1} + m_{12} S_{21} + m_{13} S_{31})^{-1} m_{13}, \\
&& U_{23} = (I_{n_2} + m_{23} S_{32} \\
&& \hskip 2em {} - (S_{21} + m_{23} S_{31}) (I_{n_1} + m_{12} S_{21}
+ m_{13} S_{31})^{-1} (m_{12} + m_{13} S_{32}))^{-1} \\
&& \hskip 4em {} \times (m_{23} - (S_{21} + m_{23} S_{31}) (I_{n_1} +
m_{12} S_{21} + m_{13} S_{31})^{-1} m_{13}).
\end{eqnarray*}

The above consideration can be directly generalised to the case of
the $\mathbb Z$-gradation of $\mathfrak{gl}(n, \mathbb C)$ which
leads to the natural representation of $n \times n$ matrices as $p
\times p$ block matrices. The corresponding equations look more and
more complicated. Nevertheless, at least for the case of constant
mappings $\lambda$ and for the case of block upper or lower
triangular
mappings $\lambda$, they can be explicitly integrated. Actually these
gradations exhaust in a sense all possible $\mathbb Z$-gradations
of the Lie algebra $\mathfrak{gl}(n, \mathbb C)$ \cite{RS97a,R98}.

\section{Multidimensional Riccati-type equations}

Let now $\lambda_i$, $i = 1, \ldots, d$ be some $\mathfrak g$-valued
functions on $\mathbb R^d$ whose standard coordinates are denoted by
$x^i$.
Consider the following system of equations for a mapping $\psi$ from
$\mathbb R^d$ to the Lie group $G$:
\begin{equation}
\partial_i \psi = \lambda_i \, \psi, \label{36}
\end{equation}
where $\partial_i = \partial/\partial x^i$.
The integrability conditions for system (\ref{36}) look as
\begin{equation}
\partial_i \lambda_j - \partial_j \lambda_i + [\lambda_i, \lambda_j]
= 0. \label{45}
\end{equation}
Similarly to the one dimensional case we obtain the following
equations for the component $\psi_{>0}$ entering the Gauss
decomposition of type (\ref{3}):
\begin{equation}
\partial_i \psi_{>0} \, \psi_{>0}^{-1} = (\psi_{>0} \, \lambda_i \,
\psi_{>0}^{-1})_{>0}.
\label{42}
\end{equation}
We call these equations {\it multidimensional Ricatti-type
equations\/}. 
The integration of equations (\ref{42}) is again reduced to the
integration of linear system (\ref{36}).

The transformation (\ref{9}), where $\chi$ is a mapping from $\mathbb
R^d$ to $G_0$, cannot be used now to get the Riccati-type equations
with $\lambda_0 = 0$. Indeed, to this end we should solve the
equations
\begin{equation}
\chi^{-1} \, \partial_i \chi  = (\lambda_i)_0. \label{46}
\end{equation}
The integrability conditions for these equations do not in general
follow from (\ref{45}). However, for the
case $(\lambda_i)_{>0} = 0$ relations (\ref{46}) are consequence of
relations (\ref{45}) and we can, with the help of transformation
(\ref{9}), reduce these equations to the case where
$(\lambda_i)_0 = 0$. Note that in the multidimensional case it is
again possible to use transformation (\ref{9}), where $\chi$ is 
some solution of the Riccati-type equations, to reduce the
equations to the case where $(\lambda_i)_{>0} = 0$.

When $\lambda_i$ are constant mappings, conditions
(\ref{45}) imply that the matrices $\lambda_i$ commute. Here, 
by a similarity transformation, we can reduce $\lambda_i$ to a
triangular form. In such case, and not only for
constant $\lambda_i$, the multidimensional Riccati-type equations can
be integrated by a procedure similar to one used in the one
dimensional case. 

As a concrete example consider the Lie group GL$(n, \mathbb C)$ with
the gradation of its Lie algebra described in section \ref{Simplest}.
Parametrising the mappings $\lambda_i$ as
\[
\lambda_i = \left( \begin{array}{cc}
A_i & B_i \\
C_i & D_i
\end{array} \right)
\]
and using for the mapping $\psi_{>0}$ parametrisation (\ref{41}) we
come to the following multidimensional Riccati-type equations:
\begin{equation}
\partial_i U = B_i - A_i U + U D_i - U C_i U.
\label{43}
\end{equation}
When $A_i = 0$ and $B_i = 0$ conditions (\ref{45}) become
as
\[
\partial_i C_j - \partial_j C_i = 0;
\]
hence, there exists a mapping $S$ such that
$C_i = \partial_i S$. Then, the general solution of equations
(\ref{43}) has the form
\[
U = (I_{n_1} + m S)^{-1} m,
\]
where $m$ is an arbitrary $n_1 \times n_2$ matrix.

\section{Generalised WZNW equations and multidimensional Toda
equations}

Consider the space $\mathbb R^{2d}$ as a differential manifold and
denote the standard coordinates on $\mathbb R^{2d}$ by $z^{-i}$,
$z^{+i}$, $i = 1, \ldots, d$. Let $\psi$ be a mapping from $\mathbb
R^{2d}$ to the Lie group $G$, which satisfies the equations
\begin{equation}
\partial_{+j} (\psi^{-1} \, \partial_{-i} \psi) = 0, \label{20}
\end{equation}
that can be equivalently rewritten as
\[
\partial_{-i}(\partial_{+j} \psi \, \psi^{-1}) = 0.
\]
Here and in what follows we use the notations $\partial_{-i} = 
\partial/\partial z^{-i}$ and $\partial_{+j} =
\partial/\partial z^{+j}$.
In accordance with \cite{GM93} we call equations (\ref{20})
the generalised WZNW equations. It is well-known that the
two dimensional Toda equations can be considered as reductions of the
WZNW equations; for a review we refer the reader to some
remarkable papers \cite{FRRTW92}, and for the affine case to
\cite{F91}. Let us show that in multidimensional
situation the appropriate reductions of the generalised WZWN
equations give the multidimensional Toda equations recently proposed
and investigated in \cite{RS97}.

It is clear that the $\mathfrak g$-valued mappings
\begin{equation}
\iota_{-i} = \psi^{-1} \, \partial_{-i} \psi, \qquad \iota_{+j} =
- \partial_{+j} \psi \, \psi^{-1} \label{37}
\end{equation}
satisfy the relations
\begin{equation}
\partial_{+j} \iota_{-i} = 0, \qquad \partial_{-i} \iota_{+j} =
0. \label{17}
\end{equation}
Moreover, the mappings $\iota_{-i}$ and $\iota_{+i}$ satisfy, by
construction, the following zero curvature conditions:
\begin{equation}
\partial_{-i} \iota_{-j} - \partial_{-j} \iota_{-i} +
[\iota_{-i}, \iota_{-j}] = 0, \qquad \partial_{+i} \iota_{+j} -
\partial_{+j} \iota_{+i} + [\iota_{+i}, \iota_{+j}] = 0.
\label{18}
\end{equation}

The reduction in question is realised by imposing on the mapping
$\psi$ the constraints
\begin{equation}
(\psi^{-1} \, \partial_{-i} \psi)_{<0} = c_{-i}, \qquad
(\partial_{+i} \psi \, \psi^{-1})_{>0} = - c_{+i}, \label{19}
\end{equation}
where $c_{-i}$ and $c_{+i}$ are some fixed mappings taking values in
the subspaces $\mathfrak g_{-1}$ and $\mathfrak g_{+1}$ respectively.
In other words, one imposes the restrictions
\[
(\iota_{-i})_{<0} = c_{-i}, \qquad (\iota_{+i})_{>0} = c_{+i}.
\]
{}From (\ref{17}) and (\ref{18}) it follows that we should consider
only the mappings $c_{-i}$ and $c_{+i}$ which satisfy the conditions
\begin{eqnarray}
& \partial_{+j} c_{-i} = 0, \qquad \partial_{-i} c_{+j} = 0,
\label{23} \\
& [c_{-i}, c_{-j}] = 0, \qquad [c_{+i}, c_{+j}] = 0. \label{24}
\end{eqnarray}
Using the Gauss decomposition (\ref{3}) we have
\begin{eqnarray*}
\psi^{-1} \, \partial_{-i} \psi = \psi_{>0}^{-1} \, \psi_0^{-1}
(\psi_{<0}^{-1} \, && \partial_{-i} \psi_{<0}) \psi_0 \, \psi_{>0} \\
&&{} + \psi_{>0}^{-1} (\psi_0^{-1} \, \partial_{-i} \psi_0) \psi_{>0}
+
\psi_{>0}^{-1} \, \partial_{-i} \psi_{>0}.
\end{eqnarray*}
Taking into account the first equality of (\ref{19}), one sees that
\begin{equation}
\psi_0^{-1} (\psi_{<0}^{-1} \partial_{-i} \psi_{<0}) \psi_0 = c_{-i}.
\label{21}
\end{equation}
Similarly one obtains the equality
\begin{eqnarray*}
\partial_{+i} \psi \, \psi^{-1} &=& \partial_{+i} \psi_{<0} \,
\psi_{<0}^{-1} \\
&+& \psi_{<0} (\partial_{+i} \psi_0 \, \psi_0^{-1}) \psi_{<0}^{-1} +
\psi_{<0} \, \psi_0 ( \partial_{+i} \psi_{>0} \, \psi_{>0})
\psi_0^{-1} \, \psi_{<0}^{-1}
\end{eqnarray*}
which implies
\begin{equation}
\psi_0 (\partial_{+i} \psi_{>0} \, \psi_{>0}^{-1}) \psi_0^{-1} =
-c_{+i}. \label{22}
\end{equation}

Let us use now the observation that the generalised WZNW equations
can be considered as the zero curvature condition for the connection
on the trivial principal fibre bundle $\mathbb R^{2d} \times G$
determined by the $\mathfrak g$-valued 1-form $\rho$ on $\mathbb
R^{2d}$ with the components
\[
\rho_{-i} = \psi^{-1} \, \partial_{-i} \psi, \qquad \rho_{+i}
= 0.
\]
After the gauge transformation of the form $\rho$ generated by the
mapping $\psi_{>0}^{-1}$ we come to the connection form $\omega$ with
the components
\[
\omega_{-i} = \psi_0^{-1} (\psi_{<0}^{-1} \, \partial_{-i} \psi_{<0})
\psi_0 + \psi_0^{-1} \, \partial_{-i} \psi_0, \qquad \omega_{+i} =
\psi_{>0} \partial_{+i} \psi_{>0}^{-1}.
\]
Since the zero curvature condition is invariant with respect to gauge
transformations, we conclude that the generalised WZNW equations are
equivalent to zero curvature condition for the form $\omega$. Using
(\ref{21}), (\ref{22}) and denoting $\psi_0$ by $\gamma$ we see that
\begin{equation}
\omega_{-i} = c_{-i} + \gamma^{-1} \, \partial_{-i} \gamma, \qquad
\omega_{+i} = \gamma^{-1} c_{+i} \gamma. \label{30}
\end{equation}
It is exactly the components of the form whose zero curvature
condition
leads to multidimensional Toda equations
\cite{RS97}\footnote{In \cite{RS97} there was considered the case
of constant $c_{-i}$ and $c_{+i}$. The generalisation to the case of
arbitrary $c_{-i}$ and $c_{+i}$ satisfying (\ref{23}) and (\ref{24})
is straightforward.} having the following explicit form
\begin{eqnarray}
& \partial_{-i} (\gamma c_{-j} \gamma^{-1}) = \partial_{-j}
(\gamma c_{-i} \gamma^{-1}), \label{26} \\
& \partial_{+j} (\gamma^{-1} \partial_{-i} \gamma) = [c_{-i},
\gamma^{-1} c_{+j} \gamma], \label{27} \\
& \partial_{+i} (\gamma^{-1} c_{+j} \gamma) = \partial_{+j}
(\gamma^{-1} c_{+i} \gamma). \label{28}
\end{eqnarray}
Thus, if a mapping $\psi$ satisfies the generalised WZNW equations
(\ref{20}) and constraints (\ref{19}), then its component $\psi_0$,
entering the Gauss decomposition (\ref{3}), satisfies
multidimensional Toda equations (\ref{26})--(\ref{28}). On the
other hand, assume that $\gamma$ is a solution of the
multidimensional Toda equations (\ref{26})--(\ref{28}); then
putting $\psi_0 = \gamma$ and choosing some $\psi_{<0}$ and
$\psi_{>0}$ which satisfy (\ref{21}) and (\ref{22}), respectively,
one can construct the solution
\[
\psi = \psi_{<0} \, \psi_0 \, \psi_{>0}
\]
of the generalised WZNW equation submitted to constraints (\ref{19}).
The explicit construction of the mappings $\psi_{<0}$ and $\psi_{>0}$
from a given solution of Toda equation for the two-dimensional case
was considered in \cite{GORS92}. Below we give the
generalisation of such construction to the multidimensional case.

First recall the procedure of obtaining the general solution to
multidimensional Toda equations \cite{RS97}. Let $\gamma_-$ and
$\gamma_+$ be some mappings from $\mathbb R^{2d}$ to $G_0$
satisfying the conditions
\[
\partial_{+i} \gamma_- = 0, \qquad \partial_{-i} \gamma_+ = 0.
\]
Consider the equations
\begin{equation}
\mu_-^{-1} \partial_{-i} \mu_- = \gamma_- c_{-i} \gamma_-^{-1},
\qquad \mu_+^{-1} \partial_{+i} \mu_+ = \gamma_+ c_{+i}
\gamma_+^{-1}, \label{25}
\end{equation}
where $\mu_-$ and $\mu_+$ obey the conditions
\[
\partial_{+i} \mu_- = 0, \qquad \partial_{-i} \mu_+ = 0.
\]
The integrability conditions for equations (\ref{25}) are
\[
\partial_{-i}(\gamma_- c_{-j} \gamma_-^{-1}) - \partial_{-j}(\gamma_-
c_{-i} \gamma_-^{-1}) = 0, \quad \partial_{+i}(\gamma_+ c_{+j}
\gamma_+^{-1}) - \partial_{+j}(\gamma_+ c_{+i} \gamma_+^{-1}) = 0.
\]
Hence, the mappings $\gamma_-$ and $\gamma_+$ cannot be arbitrary.
Suppose that the above integrability conditions are satisfied and
solve equations (\ref{25}). Consider the Gauss decomposition 
\begin{equation}
\mu_+^{-1} \mu_- = \nu_- \eta \nu_+^{-1}, \label{34}
\end{equation}
where the mapping $\nu_-$ takes values in $G_{<0}$, the mapping
$\eta$
takes values in $G_0$ and the mapping $\nu_+$ takes values in
$G_{>0}$. It can be shown \cite{RS97} that the mapping
\begin{equation}
\gamma = \gamma_+^{-1} \eta \gamma_- \label{31}
\end{equation}
satisfies the multidimensional Toda equations
(\ref{26})--(\ref{28}).

Since the manifold $\mathbb R^{2d}$ is simply connected and the
connection form $\omega$ satisfies the zero curvature condition, then
there exists a mapping $\varphi: \mathbb R^{2d} \to G$ such that
\[
\omega_{-i} = \varphi^{-1} \partial_{-i} \varphi, \qquad \omega_{+i}
= \varphi^{-1} \partial_{+i} \varphi.
\]
As it was shown in \cite{RS97}, the general form of the
mapping $\varphi$ corresponding to the solution of the
multidimensional Toda equations constructed with the help of the
above described procedure,
is
\begin{equation}
\varphi = a \mu_+ \nu_- \eta \gamma_- = a \mu_- \nu_+ \gamma_-,
\label{29}
\end{equation}
where $a$ is an arbitrary constant element of the Lie group $G$.
Using
(\ref{29}) we have
\[
\omega_{-i} = \varphi^{-1} \, \partial_{-i} \varphi = (\eta
\gamma_-)^{-1} \left(\nu_-^{-1} \, \partial_{-i} \nu_-\right) \eta
\gamma_- +
(\eta \gamma_-)^{-1} \partial_{-i} (\eta \gamma_-). 
\]
Comparing this relation with the first equality in (\ref{30}) and
taking into account (\ref{31}) we conclude that
\[
(\gamma_+^{-1} \nu_- \gamma_+)^{-1} \partial_{-i} (\gamma_+^{-1}
\nu_-
\gamma_+) = \gamma c_{-i} \gamma^{-1}.
\]
Thus we see that the general solution of  equations (\ref{21})
with $\psi_0 = \gamma$ can be written as
\begin{equation}
\psi_{<0} = \xi_-^{-1} \, \gamma_+^{-1} \, \nu_- \gamma_+, \label{32}
\end{equation}
where $\xi_-$ is an arbitrary mapping which takes values in $G_{<0}$
and satisfies the conditions
\[
\partial_{-i} \xi_- = 0.
\]
In a similar way we obtain the relation
\[
\partial_{+i} (\gamma_-^{-1} \nu_+^{-1} \gamma_-) \, (\gamma_-^{-1}
\nu_+ \gamma_-) = -\gamma^{-1} c_{+i} \gamma
\]
which implies that the general solution of equations (\ref{22}) with
$\psi_0 = \gamma$ is given by
\begin{equation}
\psi_{>0} = \gamma_-^{-1} \nu_+^{-1} \gamma_- \xi_+, \label{33}
\end{equation}
where $\xi_+$ is an arbitrary mapping which takes values in $G_{>0}$
and satisfies the
conditions
\[
\partial_{+i} \xi_+ = 0.
\]
Using relations (\ref{32}) and (\ref{33}) we come to the following
representation for the solution of the generalised WZNW equations
corresponding to the solution of the multidimensional Toda equations
$\psi_0 = \gamma$:
\[
\psi = \psi_{<0} \, \psi_0 \, \psi_{>0} = \xi_-^{-1} \gamma_+^{-1}
\nu_- \eta \nu_+^{-1} \gamma_- \xi_+.
\]
Due to relation (\ref{34}) this representation is equivalent to
\[
\psi = \xi_-^{-1} \gamma_+^{-1} \mu_+^{-1} \mu_- \gamma_- \xi_+.
\]
In the next section we use this representation to construct some
integrable classes of the multidimensional Riccati-type equations.

\section{Multidimensional Toda systems and Riccati-type \kern 1.em\
equations}

Let $\lambda_{-i}$ and $\lambda_{+i}$, $i = 1, \ldots, d$, be some
fixed mappings from the manifold $\mathbb R^{2d}$ to the Lie algebra
$\mathfrak g$ which satisfy conditions
\begin{equation}
\partial_{+j} \lambda_{-i} = 0, \qquad \partial_{-j} \lambda_{+i} =
0. \label{51}
\end{equation}
Consider the system of equations
\begin{equation}
\partial_{-i} \psi = \psi \, \lambda_{-i}, \qquad \partial_{+i}
\psi = - \lambda_{+i} \, \psi, \label{35}
\end{equation}
where $\psi$ is a mapping from $\mathbb R^{2d}$ to the Lie group $G$.
The integrability conditions for this system is given by
\begin{equation}
\partial_{-i} \lambda_{-j} - \partial_{-j} \lambda_{-i} +
[\lambda_{-i}, \lambda_{-j}] = 0, \quad \partial_{+i} \lambda_{+j} -
\partial_{+j} \lambda_{+i} + [\lambda_{+i}, \lambda_{+j}] = 0.
\label{52}
\end{equation}
It is clear that the mapping $\psi$ satisfies the generalised WZNW
equations. Hence we can treat system (\ref{35}) with the mappings
$\lambda_{-i}$ and $\lambda_{+i}$ satisfying (\ref{51}) and
(\ref{52}), as a reduction of the generalised WZNW equations similar
to the reduction considered in the previous section. The difference
is that in the previous section we fixed only the components
$(\iota_{-i})_{<0}$ and $(\iota_{+i})_{>0}$ of the the mappings
$\iota_{-i}$ and $\iota_{+i}$ and did it in a quite special way,
but here we fix the mappings $\iota_{-i}$ and $\iota_{+i}$
completely. It is easy to show that if the mapping $\psi$
satisfies equations (\ref{35}) then the mappings $\psi_{<0}^{-1}$ and
$\psi_{>0}$ satisfy the multidimensional Riccati-type equations
\begin{eqnarray}
&\partial_{+i} \psi_{<0}^{-1} \, \psi_{<0}= (\psi_{<0}^{-1} \,
\lambda_{+i} \, \psi_{<0})_{<0}, \label{39} \\
&\partial_{-i} \psi_{>0} \, \psi_{>0}^{-1} = (\psi_{>0} \,
\lambda_{-i} \, \psi_{>0}^{-1})_{>0}. \label{40}
\end{eqnarray}
Equations (\ref{35}) are multidimensional generalisation of the
so-called associated Redheffer--Reid system \cite{Red56Rei59}. The
investigation of that system is very useful for studying
one dimensional Riccati and matrix Riccati equations, see for
example \cite{ZI73}. We believe that our generalisation  also
play a significant role for the multidimensional Riccati-type
equations. As a first application of such systems let us give a
construction of some integrable class of the multidimensional
Riccati-type equations.

Suppose now that the mappings $\lambda_{-i}$ and $\lambda_{+i}$
are that
\begin{equation}
(\lambda_{-i})_{<0} = c_{-i}, \qquad (\lambda_{+i})_{<0} = c_{+i}
\label{53}
\end{equation}
with the mappings $c_{-i}$ and $c_{+i}$ taking values in $\mathfrak
g_{-1}$ and $\mathfrak g_{+1}$, respectively, and submitted to
conditions (\ref{23}) and (\ref{24}). In this case the mapping
$\gamma = \psi_0$ satisfies the multidimensional Toda equations
(\ref{26})--(\ref{28}). On the other hand, if we have a solution
$\gamma$ of equations (\ref{26})--(\ref{28}), then using results of
the previous section we can find the general solution to equations
(\ref{21}) and (\ref{22}), and construct the mapping $\psi$ which
satisfies the generalised WZNW equations and constraints (\ref{19}).
This mapping, via equalities (\ref{35}), generates some mappings
$\lambda_{-i}$ and $\lambda_{+i}$ certainly satisfying constraints
(\ref{53}). Actually if we have the general solution to
multidimensional Toda equations (\ref{26})--(\ref{28}), then we get
in this way the general form of the mappings $\lambda_{-i}$ and
$\lambda_{+i}$ which satisfy the integrability conditions (\ref{52})
and constraints (\ref{53}).
Moreover, we have here the general solution to the multidimensional
Riccati equations (\ref{39}) and (\ref{40}).

The explicit form of the mappings $\lambda_{-i}$ and $\lambda_{+i}$
obtained with the help of the above described procedure is
\begin{eqnarray*}
& \lambda_{-i} = \xi_+^{-1} \, c_{-i} \, \xi_+ + \xi_+^{-1}
(\gamma_-^{-1} \, \partial_{-i} \gamma_-) \xi_+ + \xi_+^{-1} \,
\partial_{-i} \xi_+, \\
& \lambda_{+i} = \xi_-^{-1} \, \partial_{+i} \xi_- + \xi_-^{-1}
(\gamma_+^{-1} \, \partial_{+i} \gamma_+) \xi_- + \xi_-^{-1} \,
c_{+i} \, \xi_-,
\end{eqnarray*}
and the corresponding solutions of equations (\ref{39}) and
(\ref{40}) are given by (\ref{32}) and (\ref{33}).

Consider the Lie group GL$(n, \mathbb C)$ and the $\mathbb
Z$-gradation of the Lie algebra $\mathfrak{gl}(n, \mathbb C)$
discussed in section \ref{Simplest}. Parametrise the mappings
$\gamma_\mp$ as
\[
\gamma_\mp = \left( \begin{array}{cc}
\beta_{\mp 1} & 0 \\
0 & \beta_{\mp 2}
\end{array} \right).
\]
The general form of the mappings $c_{\mp i}$ is
\[
c_{-i} = \left( \begin{array}{cc}
0 & 0 \\
X_{-i} & 0
\end{array} \right), \qquad
c_{+i} = \left( \begin{array}{cc}
0 & X_{+i} \\
0 & 0
\end{array} \right),
\]
where the mappings $X_{-i}$ and $X_{+i}$ are arbitrary. The
integrability conditions of equations (\ref{25}) have now the form
\begin{eqnarray}
&\partial_{-i}(\beta_{-2} X_{-j} \beta_{-1}^{-1}) -
\partial_{-j}(\beta_{-2} X_{-i} \beta_{-1}^{-1}) = 0, \label{47} \\
&\partial_{+i}(\beta_{+1} X_{+j} \beta_{+2}^{-1}) -
\partial_{-j}(\beta_{+1} X_{-i} \beta_{+2}^{-1}) = 0. \label{48}
\end{eqnarray}
For the mappings 
\[
\lambda_{\mp i} = \left( \begin{array}{cc}
A_{\mp i} & B_{\mp i} \\
C_{\mp i} & D_{\mp i}
\end{array} \right)
\]
we obtain
\begin{eqnarray*}
&& A_{-i} = \beta_{-1}^{-1} \partial_{-i} \beta_{-1} - (\xi_+)_{12}
X_{-i}, \\
&& B_{-i} = \beta_{-1}^{-1} \partial_{-i} \beta_{-1} (\xi_+)_{12} -
(\xi_+)_{12} \beta_{-2}^{-1} \partial_{-i} \beta_{-2} - (\xi_+)_{12}
X_{-i} (\xi_+)_{12} + \partial_{-i} (\xi_+)_{12}, \\
&& C_{-i} = X_{-i}, \quad D_{-i} = \beta_{-2}^{-1} \partial_{-i}
\beta_{-2} + X_{-i} (\xi_+)_{12}, \\
&& A_{+i} = \beta_{+1}^{-1} \partial_{+i} \beta_{+1} + X_{+i}
(\xi_-)_{21}, \quad B_{+i} = X_{+i}, \\
&& C_{+i} = \beta_{+2}^{-1} \partial_{+i} \beta_{+2} (\xi_-)_{21} -
(\xi_-)_{21} \beta_{+1}^{-1} \partial_{+i} \beta_{+1} - (\xi_-)_{21}
X_{+i} (\xi_-)_{21} + \partial_{+i} (\xi_-)_{21}, \\
&& D_{+i} = \beta_{+2} \partial_{+i} \beta_{+2} - (\xi_-)_{21}
X_{+i},
\end{eqnarray*}
where $(\xi_+)_{12}$ and $(\xi_-)_{21}$ are the nontrivial blocks of
the mappings $\xi_+$ and $\xi_-$.

In order to solve equations (\ref{39}) and (\ref{40}) one considers
first equations (\ref{25}). Next, one uses the Gauss decomposition
(\ref{34}) for finding the mappings $\nu_+^{-1}$ and $\nu_-$. In the
case under consideration 
\begin{eqnarray*}
&& \nu_+^{-1} = \left( \begin{array}{cc}
I_{n_1} & - (I_{n_1} - (\mu_+)_{12} (\mu_-)_{21})^{-1} (\mu_+)_{12}
\\
0 & I_{n_2}
\end{array} \right), \\
&& \nu_- = \left( \begin{array}{cc}
I_{n_1} & 0 \\
(\mu_-)_{21} (I_{n_1} - (\mu_+)_{12} (\mu_-)_{21})^{-1} & I_{n_2}
\end{array} \right).
\end{eqnarray*}
Finally, using (\ref{33}) and (\ref{32}) one arrives at the
following
expressions for nontrivial blocks $(\psi_{>0})_{12} = U_-$ and
$(\psi_{<0})_{21} = U_+$ of the mappings $\psi_{>0}$ and $\psi_{<0}$:
\begin{eqnarray*}
&& U_- = (\xi_+)_{12} - \beta_{-1}^{-1} (I_{n_1} - (\mu_+)_{12}
(\mu_-)_{21})^{-1} (\mu_+)_{12} \beta_{-2}, \\
&& U_+ = (\xi_-)_{21} + \beta_{+2}^{-1} (\mu_-)_{21} (I_{n_1} -
(\mu_+)_{12} (\mu_-)_{21})^{-1} \beta_{+1}.
\end{eqnarray*}
It is clear that the dependence of $U_-$ and $U_+$ on $z^{+i}$
and $z^{-i}$, respectively, is parametric, and the general solution
of the equations can be written as
\begin{eqnarray}
&& U_- = (\xi_+)_{12} - \beta_{-1}^{-1} (I_{n_1} - m_-
(\mu_-)_{21})^{-1} m_- \beta_{-2}, \label{49} \\
&& U_+ = (\xi_-)_{21} + \beta_{+2}^{-1} m_+ (I_{n_1} -
(\mu_+)_{12} m_+)^{-1} \beta_{+1}, \label{50}
\end{eqnarray}
where $m_-$ and $m_+$ are arbitrary constant matrices of dimensions
$n_1 \times n_2$ and $n_2 \times n_1$ respectively.

We have said nothing yet about solving integrability conditions
(\ref{47}) and (\ref{48}). In the general case the solution to these
equations is not known. However, they can be solved in some
particular cases. For example, let $n = d + 1$, $n_1 = d$ and $n_2 =
1$, and let the mappings $X_{\mp i}$ be defined by the relations
\[
(X_{-i})_{1j} = \delta_{ij}, \qquad (X_{+i})_{j1} = \delta_{ij}.
\]
In this case the general solution \cite{RS97} of integrability
conditions (\ref{47}) and (\ref{48}) is 
\begin{eqnarray*}
(\beta_{-1}^{-1})_{ij} = F_- \partial_{-i} H_{-j}, \qquad
\beta_{-2}^{-1} = F_-, \\
(\beta_{+1})_{ij} = F_+ \partial_{+j} H_{+i}, \qquad \beta_{+2} =
F_+,
\end{eqnarray*}
where $F_\mp$ and $H_{\mp i}$ are arbitrary functions depending on
the coordinates $z^{\mp i}$. 
For the blocks $(\mu_-)_{21}$ and $(\mu_+)_{12}$ one has
\[
(\mu_-)_{21} = H_-, \qquad (\mu_+)_{21} = H_+,
\]
where $H_-$ and $H_+$ are $1 \times d$ and $d \times 1$ matrices
formed by the functions $H_{-i}$ and $H_{+i}$ respectively. Now using
the evident notations we can write expressions (\ref{49}) and
(\ref{50}) as
\[
U_{-i} = \xi_{+i} + \partial_{-i} \log (1 - H_- m_-), \qquad U_{+i} =
\xi_{-i} - \partial_{+i} \log (1 - m_+ H_+).
\]

\begin{center}
{\large\bf Acknowledgements}
\end{center}

The authors are indebted to A.~M.~Bloch and A.~K.~Common who
acquainted
us with their studies related to the matrix ordinary differential
Riccati equation. One of the authors (M.~V.~S.) is grateful to
J.--L.~Gervais for useful discussions; he also wishes to acknowledge
the warm hospitality of the Instituto de F\'\i sica Te\'orica,
Universidade Estadual Paulista, S\~ao Paulo, Brazil, and the
financial support from FAPESP during his stay there in March--July
1998. The research program of A.~V.~R. and M.~V.~S. is supported in
part by the Russian Foundation for Basic Research under grant \#
98--01--00015 and by INTAS grant \# 96-690; and that of L.~A.~F.,
J.~F.~G. and A.~H.~Z. is partially supported by CNPq-Brazil.

\end{document}